# Introductory Quantum Physics Courses using a LabVIEW multimedia module

Ismael Orquín[1,*], Miguel-Ángel García-March[1], Pedro Fernández de Córdoba[1],
Javier F. Urcheguía[2], and Juan A. Monsoriu[2]

[1]Department of Applied Mathematics, Polytechnical University of Valencia, 46022 Valencia, Spain

[2]Department of Applied Physics, Polytechnical University of Valencia, 46022 Valencia, Spain

[*]isorser@doctor.upv.es

**ABSTRACT**

We present the development of a LabVIEW multimedia module for introductory Quantum Physics courses and our experience in the use of this application as an educational tool in learning methodologies. The program solves the Time Dependent Schrödinger Equation for arbitrary potentials. We describe the numerical method used for solving this equation, as well as some mathematical tools employed to reduce the calculation time and to obtain more accurate results. As an illustration, we present the evolution of a wave packet for three different potentials: the repulsive barrier potential, the repulsive step potential, and the harmonic oscillator. This application has been successfully integrated in the learning strategies of the course *Quantum Physics for Engineering* at the Polytechnic University of Valencia, Spain.

**Keywords:** Finite Difference Method, LabVIEW, Schrödinger Equation, Digital Simulation.

# 1. Introduction.

Recently Information and Communication Technologies (there after ICTs) have changed the conception of the teaching process both in the classroom and in the theoretical teaching approaches. Traditional content-oriented teaching approaches are being replaced by student-oriented ones. In agreement with the principles of constructivism, ICTs benefit from permanent interactivity [1]. Particularly, learning strategies based on digital simulations have demonstrated to provide additional advantages [2,3]. However, it should be emphasized that pedagogical effectiveness is related to the complete integration of the simulation in the overall instruction plan [4,5].

The introduction of ICTs for teaching permits to present the contents of the course from different points of view. For example, computer simulations allow the development of virtual experiments in which the students can decide in situ the correct values to obtain and visualize certain physical phenomena [6]. In our opinion, this learning strategy is much better for students both in terms of understanding and visualizing than just passively attending lectures, because it permits a closer interaction between the students and the contents of the course.

This learning methodology has been applied in the course *Quantum Physics for Engineering* at the Polytechnic University of Valencia (PUV), Spain, where it was incorporated to the study program of the Higher Technical School of Industrial Engineering in 2002. The instruction plan of this course includes a wide variety of learning tools, with the purpose of making the learning process easier for the students, e.g., solving practical problems in the classroom with the help of the teacher, playing educational videos about physical phenomena and using digital simulation programs, like the one reported here. All these tools serve as complement to the traditional lectures where the teacher explains the theoretical contents of the course with help of the text-book [7].

As part of our teaching activities we have developed interactive digital simulation programs that illustrate physical concepts as a complement to the theoretical explanations. In this paper we present a LabVIEW multimedia application for introductory Quantum Physics courses. LabVIEW is a graphical programming environment using icons quite unlike the text-based programming with conventional languages such as C or FORTRAN. We can find in the literature examples of pedagogical LabVIEW applications for Physics and Engineering [8,9]. Our LabVIEW stand-alone executable program solves the Time Dependent Schrödinger Equation (TDSE) by a finite difference technique. Implicit Crank Nicolson scheme is normally employed to solve this equation because of its unconditional stability. LabVIEW provides interesting features to develop Graphical User Interfaces (GUI) characterized by their friendly appearance so that they are easy to use as well as very attractive for the user. With this program we are able to simulate the dynamics of a particle under the effect of different potentials.

This paper is organized as follows. In Section 2 we revise the solution for the TDSE as well as the numerical techniques employed to obtain the solution. In Section 3 we discuss the implementation of the finite difference scheme in LabVIEW and present some numerical results of the dynamics of a wave packet at different potentials. We describe some tools that reduce calculation time. Finally, Section 4 summarizes our main results.

**2. Numerical Formulation.**

The problem to solve corresponds to the one-dimensional dynamics of a quantum particle at a given potential. Particle movement is described by the Time Dependent Schrödinger Equation [7] given by

$$-\frac{\hbar^2}{2m}\cdot\frac{\partial^2}{\partial x^2}\Psi(x,t)+V(x)\cdot\Psi(x,t)=i\hbar\frac{\partial}{\partial t}\Psi(x,t), \qquad (1)$$

where $\Psi(x,t)$ is the wave function of a particle at position $x$ at time $t$, and $m$ is the particle mass. $\int_{x_0}^{x_1}|\Psi(x,t)|^2 dx$ is the probability of locating the particle between the positions $x_0$ and $x_1$ at time $t$. The solution of equation (1) provides the time evolution of the wavefunction under the influence of the potential defined by the function $V(x)$. To solve this differential equation we need an initial value for the function so as to permit the analysis of its time evolution. Thus, we will be dealing with a problem of initial values. The initial gaussian wavefunction takes the usual form

$$\Psi(x,t=0)=\frac{1}{\sqrt{\sigma\sqrt{2\pi}}}\cdot\exp(ikx)\cdot\exp\left(-\frac{(x-x_0)^2}{4\sigma^2}\right), \qquad (2)$$

where $x_0$ and $\sigma$ are the initial position and width, respectively, of the wavefunction and $k$ is the wavevector of the particle. Note that the wavefunction is normalized to $\int_{-\infty}^{+\infty}|\Psi(x,t_0)|^2 dx=1$.

Our code works with the dimensionless time-dependent Schrödinger equation given by

$$\left(-\frac{\partial^2}{\partial x'^2}+V'(x')\right)\cdot\Psi(x',t')=i\cdot\frac{\partial\Psi(x',t')}{\partial t'}, \qquad (3)$$

which is obtained from Eq. (1) by using the dimensionless variables, $x'$ and $t'$ defined as

$$t'=\frac{m\cdot c^2}{2\cdot\hbar}t \qquad \text{and} \qquad x'=\frac{m\cdot c}{\hbar}x. \qquad (4)$$

These new variables give

$$V'(x')=\frac{2}{m\cdot c^2}V(x'). \qquad (5)$$

The implicit Crank-Nicolson scheme used to solve Eq. (3) consists of the discretization of the continuous spatial and temporal domain where the wavefunction, $\Psi(x',t')$, is defined. We choose both spatial and temporal steps of discretization $\Delta x'$ and $\Delta t'$ so that the discrete variables become $x'_j = x'_0+j\Delta x'$ and $t'^n = t'_0+n\Delta t'$, where $j$ and $n$ are integers. The approximation to the solution at time step $t'^n$ and at spatial step $x'_j$ is $\Psi_j^n = \Psi(x_j,t_n)$ and the potential at this position is $V'_j= V'(x_j)$. The finite difference operators that approximate the derivatives appearing in the TDSE are

$$\frac{\partial \Psi}{\partial t'} \approx \frac{\Psi_j^{n+1} - \Psi_j^n}{\Delta t'} , \qquad (6a)$$

$$\frac{\partial^2 \Psi}{\partial x'^2} \approx \frac{1}{2}\left(\frac{\Psi_{j-1}^{n+1} - 2\Psi_j^{n+1} + \Psi_{j+1}^{n+1}}{\Delta x'^2} + \frac{\Psi_{j-1}^n - 2\Psi_j^n + \Psi_{j+1}^n}{\Delta x'^2}\right). \qquad (6b)$$

Then the difference equation becomes

$$i \cdot \frac{\Psi_j^{n+1} - \Psi_j^n}{\Delta t'} = -\frac{1}{2}\cdot\left(\frac{\Psi_{j-1}^{n+1} - 2\Psi_j^{n+1} + \Psi_{j+1}^{n+1}}{\Delta x'^2} + \frac{\Psi_{j-1}^n - 2\Psi_j^n + \Psi_{j+1}^n}{\Delta x'^2}\right) + V'_j \frac{1}{2}\left(\Psi_j^{n+1} + \Psi_j^n\right). \qquad (7)$$

Keeping the terms corresponding to the unknown time interval $n+1$ to the left side and the known variables at $n$ time step to the right, we obtain,

$$-\Psi_{j-1}^{n+1} + \left(2 + \frac{2\cdot\Delta x'^2}{i\cdot\Delta t'} + V'_j \Delta x'^2\right)\Psi_j^{n+1} - \Psi_{j+1}^{n+1} =$$

$$= \Psi_{j-1}^n + \left(-2 + \frac{2\cdot\Delta x'^2}{i\cdot\Delta t'} - V'_j \Delta x'^2\right)\Psi_j^n + \Psi_{j+1}^n \qquad (8)$$

Using a more compact notation defined by

$$\alpha_j = 2 + \frac{2\cdot x'^2}{i\cdot t'} + V_j \Delta x'^2 \qquad \beta_j = -2 + \frac{2\cdot x'^2}{i\cdot t'} - V_j \Delta x'^2 , \qquad (9)$$

Eq. (8) can be written as

$$-\Psi_{j-1}^{n+1} + \alpha_j \Psi_j^{n+1} - \Psi_{j+1}^{n+1} = \Psi_{j-1}^n + \beta_j \Psi_j^n + \Psi_{j+1}^n. \qquad (10)$$

This scheme yields a linear system of equations for each time step with as many equations as spatial steps to solve the problem. We must solve $\Psi_j^n$ for each value of $j$ and $n$. The values of the function at each extreme of the spatial domain are given by boundary conditions. Usually the wave function is zero at the boundary of the spatial domain when the solution does not arrive at the boundary and the proper dynamics of the particle at the particular potential can be studied [10]. In order to solve the linear system of equations given by Eq. (10), it can be writen in matrix notation as

$$\begin{pmatrix} \alpha_1 & -1 & 0 & \dots \\ -1 & \alpha_2 & -1 & \dots \\ 0 & -1 & \alpha_3 & \dots \\ \dots & \dots & \dots & \dots \end{pmatrix} \cdot \begin{pmatrix} \Psi_1^{n+1} \\ \Psi_2^{n+1} \\ \Psi_3^{n+1} \\ \dots \end{pmatrix} = \begin{pmatrix} \beta_1 & 1 & 0 & \dots \\ 1 & \beta_2 & 1 & \dots \\ 0 & 1 & \beta_3 & \dots \\ \dots & \dots & \dots & \dots \end{pmatrix} \cdot \begin{pmatrix} \Psi_1^n \\ \Psi_2^n \\ \Psi_3^n \\ \dots \end{pmatrix}. \qquad (11)$$

In (11) we know the wavefunction at the time step $n$, $\Psi_j^n$, and the solution of this linear system of equations provides the wavefunction at the next time step, $\Psi_j^{n+1}$. The Crank Nicolson scheme is chosen because it is unconditionally stable regardless of the step size used to discretize the equation. The finite difference method will be stable although it may not be exact if a big step size is chosen. Besides, the square of the absolute value of the function must remain constant throughout the calculation process, which is a very important characteristic of the problem we are dealing with.

The initial condition for the wave packet is calculated in a subroutine which is called from the main program. This routine is used to generate the initial values of the wave packet for each value of the spatial variable. So it is called as many times as spatial steps for a given time step.

**3. Presentation of results.**

We have chosen LabVIEW in order to implement the previous formulation for the teaching purpose. LabVIEW is an object-oriented language for virtual instruments programming. ICTs allow now to simulate laboratory tools by using a personal computer so that we don't have to buy the expensive instruments used in a laboratory for signal analysis, for example. This kind of programs that simulate the behaviour of complex devices are called virtual instruments and LabVIEW is one of the languages allowing virtual instrument programming.

Due to its flexibility and capability for doing calculations in an easy manner, LabVIEW is a useful programming language to obtain very friendly graphical interfaces, as the one presented here. Additionally, LabVIEW RunTime Engine makes it possible to obtain a stand-alone application that can be installed in any personal computer which has this module installed. In this way there is no need to install the complete LabVIEW environment. It also permits the programmer to choose the controls, i.e., the controls are those variables that the user can change from the Control Panel. The indicators display the user the results of the calculations of our program; they can be distinguished from the controls because indicators can not be modified by the user. All these reasons justify the selection of LabVIEW as a very interesting simulation tool of physical phenomena.

With respect its programming language, an object-oriented language is much closer to intuition than any other language. For example, in Fig. 1 we show the code that represents the kernel of the finite difference calculation. The upper diagram represents the calculation of the right hand side of equation (9). This term correspond to independent vector *b* in the linear system of equations given by *Ax=b*. Observe that the coefficient matrix is tridiagonal. Then, we proceed to calculate the *LU* decomposition of matrix *A* to solve the system of equations in a more effective way by forward and backward substitution of the two underlying triangular systems of equations given by *Ly=b* and *Ux=y*. This is represented in the lower diagram of figure 1. The calculation makes the resolution of the system of equations (11) much quicker.

The *Front Panel* of the application is shown in Fig. 2. The user has to determine the properties of the wave packet at *t* = 0, the characteristics of the potential to simulate and the size of the steps used to discretize the equation in both variables (temporal and spatial). The wavefunction at *t* = 0 is a gaussian wave packet scaled to one. The initial position of the packet, its central momentum and its standard deviation, must be defined by the user. Different values have to be defined for each potential. For example, the user must define the frequency of the harmonic oscillator potential or the position and height of the energy barrier when simulating the barrier potential. For the characterization of the finite difference method, the user has to define the size of the steps used to discretize the equation are required as well as the number of intervals necessary to discetize the two-dimensional domain. It is important to remember that the selection of this parameter i.e., (step size) will determine the accuracy of our numerical approximation to the solution.

The tool presented here has been applied in the course to study the interaction of a wave packet with arbitrary potentials. First, the students can see how the bound states of each potential evolve with no changes in their shape because they are solutions to the Time Independent Schrodinger Equation. The students can also study other additional aspects of the dynamics of a quantum particle relative to particle interaction with different potentials. The program permits to save the data into a text file in which we can store the evolution of the position and momentum expectation values, as well as the transmission and reflection coefficients. Figs. 3, 4, and 5, present the results obtained for different potentials.

- ***Repulsive Barrier Potential.*** As an illustration in Fig. 3 we present three different frames taken from the simulation of the dynamics of a particle that collides against a repulsive barrier. We can see the transmission of the wave packet through the barrier.
- ***Repulsive Step Potential.*** Fig. 4 shows the simulation corresponding to a particle colliding against a repulsive step potential. We can see that the wavefunction is likely to stay inside the repulsive potential temporarily. In fact, this phenomenon should not get confused with the transmission of the particle, which in fact, does not occur at this potential.
- ***Harmonic Oscillator***. Fig. 5 shows the evolution of a wave packet under the effect of a quadratic harmonic oscillator. The wave packet experiences a periodic movement where the period depends on the frequency of the harmonic oscillator.

**4. Conclusions.**

In this paper we have presented a new pedagogical tool to complement the teaching of Quantum Physics in the studies of Engineering. It is a simulation program based on LabVIEW. Students can use the program at any moment of their instruction and as many times as required. The use of this tool in the Lab sessions helps to improve the student's understanding of quantum phenomena. From our point of view, this simulation program has two principal advantages over more traditional teaching materials: a) the users are able to study the influence of the different variables of the problem without having to solve the entire problem every time, and b) the visualization of the process helps to understand the quantum phenomena. With respect to the learning strategies, the program allows the use of different integrated techniques that help to reach a greater variety of students. Furthermore, this simulation program exports calculation data so that the students can write a short report describing the simulations they have performed explaining the results obtained.

**Acknowledgment**

The authors are thankful to Professor Sarira Sahu from Instituto de Ciencias Nucleares, Universidad Nacional Autonoma de Mexico, Mexico, for his valuable comments and suggestions. I. Orquín and M.A. García-March gratefully acknowledge the grant PID 13042-C from Instituto de Ciencias de la Educación, PUV. This work has been supported by Ministerio de Ciencia y Tecnología, Spainh Government, under research project TIC2002-04527-C02-02.

**REFERENCES**

[1] T. Duffy and K. Jonassen, *Constructivism and the technology of instruction*, Lawrence Erlbaum Associates, Hilsdale, New Jersey, 1992.

[2] F. Esquembre, "Computers in physics education," *Comput. Phys. Commun.*, Vol. 147, 2002, pp. 13-18.

[3] D. J. Grayson and L.C. McDermott, "Use of computer for research on student thinking in physics", *Am. J. Phys.*, Vol. 64, 1996, pp.557-565.

[4] D. Hestenes, "Who needs physics education research?," *Am. J. Phys.*, Vol. 66, 1998, pp. 465-467.

[5] R.N. Steinberg, "Computers in teaching science: To simulate or not to simulate," *Am. J. Phys*, Vol. 68, 2000, pp. S37-S41.

[6] A. Vidaurre, J. Riera, M.H. Jiménez, and J.A. Monsoriu, "Contribution of simulation in visualizing physics processes," *Comput. Appl. Eng. Educ.*, Vol. 10, 2002, pp. 45-49.

[7] P. Fernández de Córdoba and J.F. Urchueguía, *Fundamentals of Quantum Physics for Engineering*, Ed. Polytechnic University of Valencia, Spain, 2004.

[8] P.J. Moriarty, B.L. Gallagher, C.J. Mellor, and R.R. Baines, "Graphical computing in the undergraduate laboratory: Teaching and interfacing with LabVIEW," *Am. J. Phys*, Vol. 71, 2003, pp. 1062-1074.

[9] S. Sanz, M.F. Iskander, and L. Yu, "Development of an Interactive Multimedia Module on Antenna Theory and Design," *Comput. Appl. Eng. Educ.*, Vol. 8, 2000, pp. 11-17.

[10] P. Fernández de Córdoba, J.F. Urchueguía, J.A. Monsoriu, and I. Orquín, "Computer simulations in Quantum Physics," *International Conference on Engineering Education*, Valencia, Spain (2003), Proceedings in CD-ROM.

**FIGURE CAPTIONS**

**Figure 1.** Resolution of Eq. (11) with a LabVIEW programming.

**Figure 2.** Front Panel of the GUI developed with LabVIEW.

**Figure 3.** Collision against a repulsive barrier. Time steps: 500, 1000, 2000.

**Figure 4.** Collision against a repulsive step. Time steps :1000, 1700, 2000.

**Figure 5.** Harmonic Oscillator Potential. Time steps: 10, 1100, 3400.

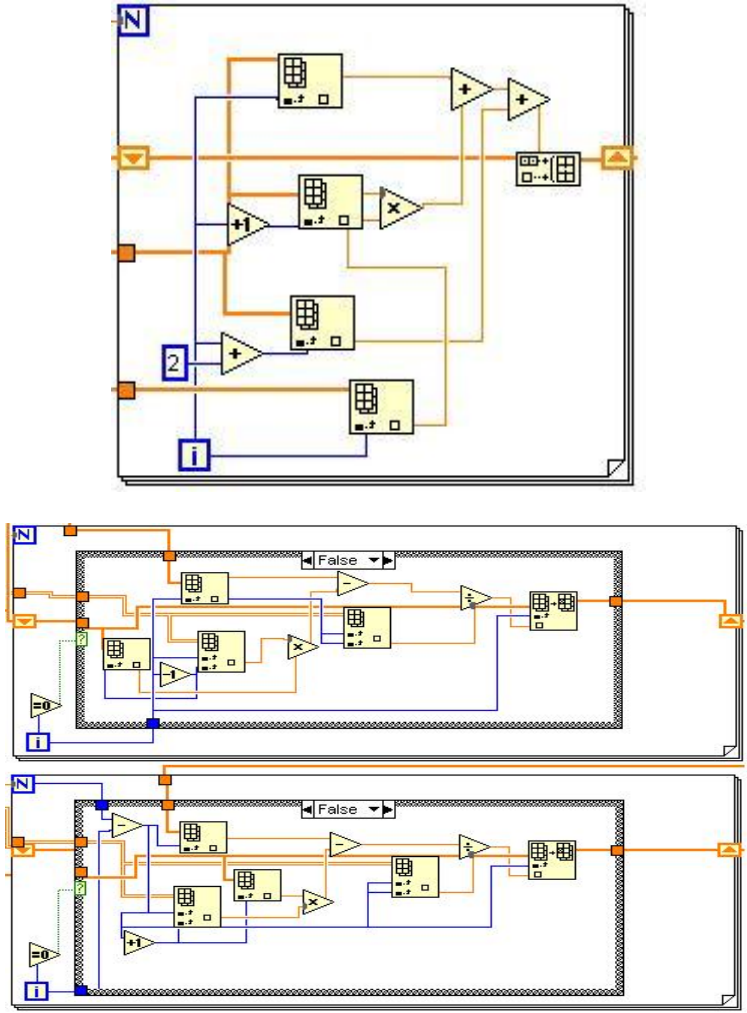

Figure 1
(Orquín *et al.*)

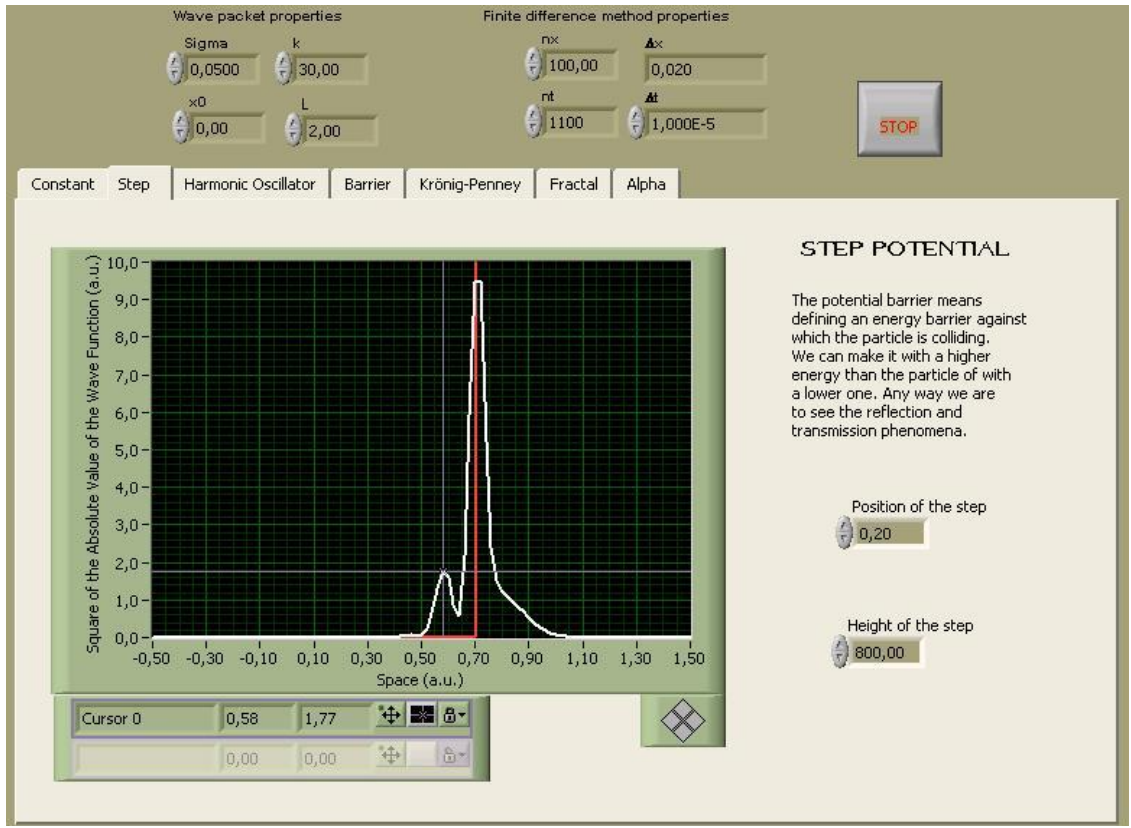

Figure 2
(Orquín *et al.*)

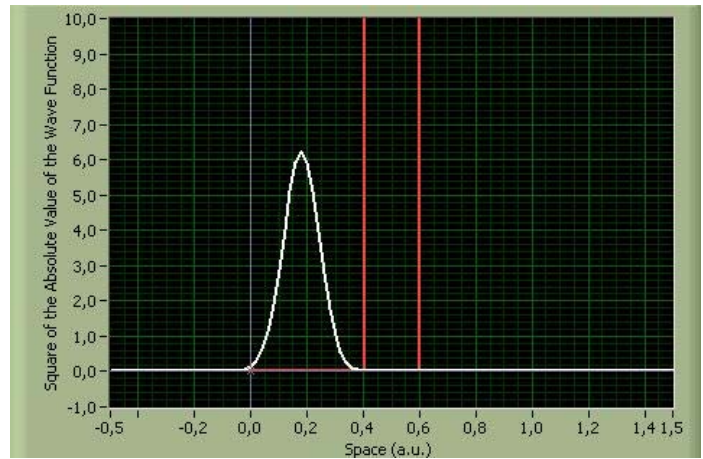

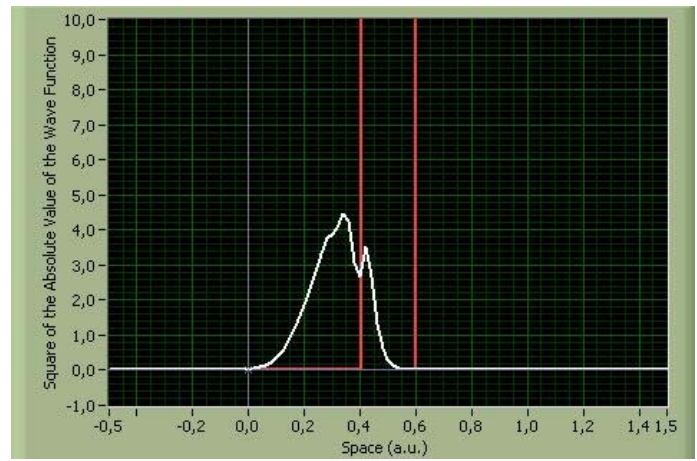

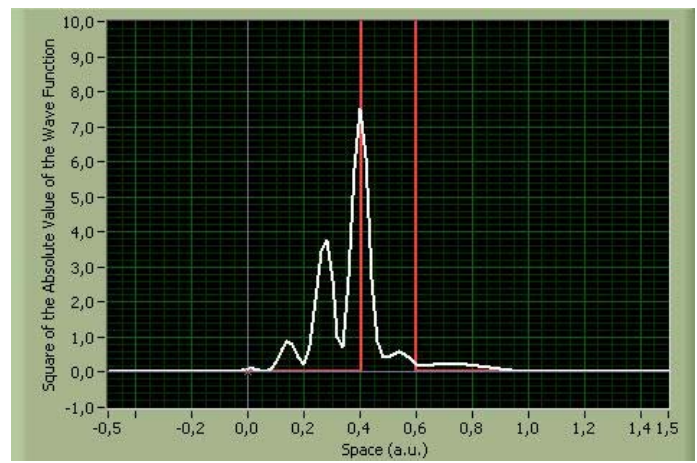

Figure 3
(Orquín *et al.*)

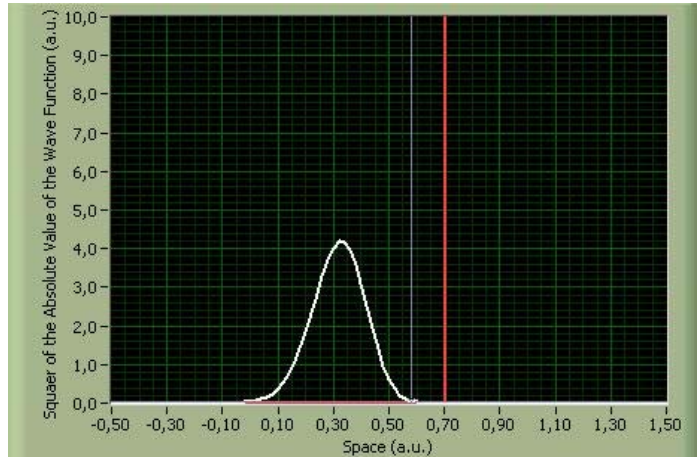

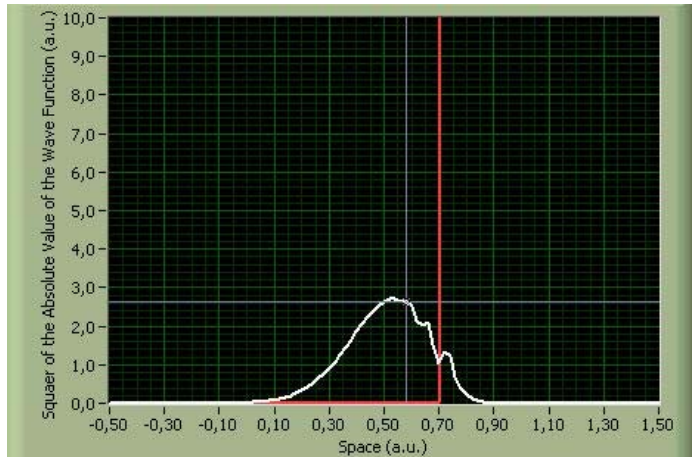

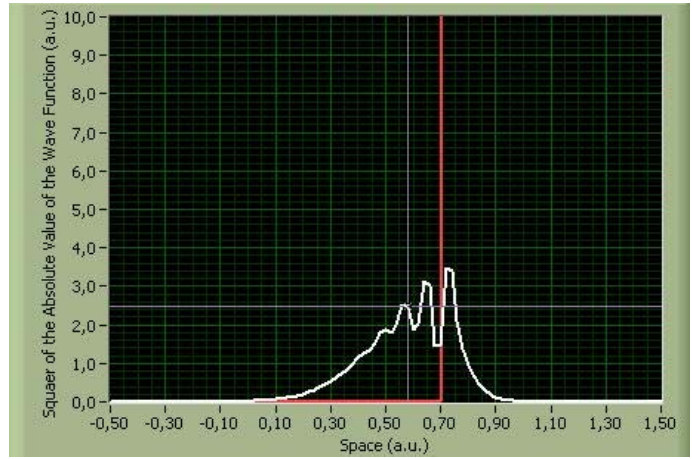

Figure 4
(Orquín *et al.*)

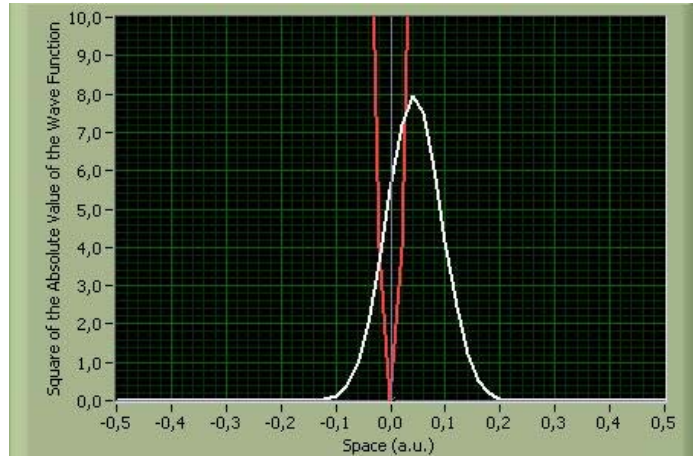

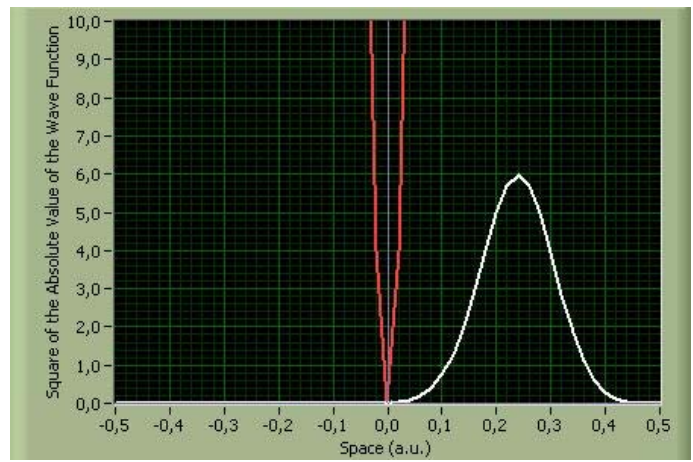

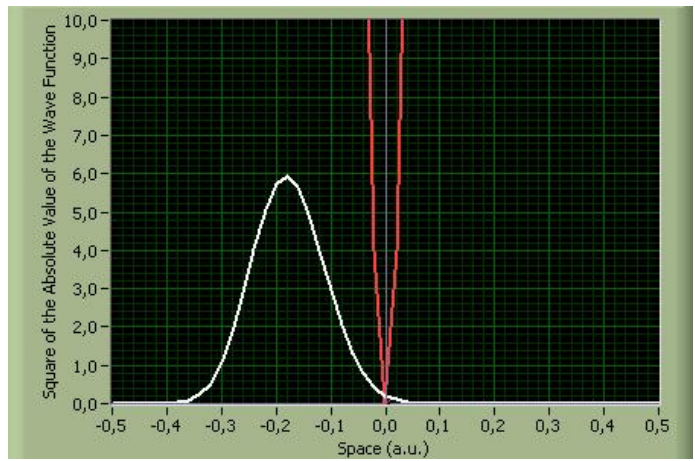

Figure 5
(Orquín *et al.*)

12